\documentclass{WileyMSP-template}

\usepackage{CJKutf8}
\usepackage{latexsym}
\usepackage{amstext}
\usepackage{amsfonts}
\usepackage{dsfont}
\usepackage{color}
\usepackage{amssymb}
\usepackage{amsmath}
\usepackage{relsize}
\usepackage{amsxtra}
\usepackage{verbatim}
\usepackage{hyperref}
\usepackage{graphicx}

\usepackage{caption}
\usepackage[labelfont=bf]{caption}
\usepackage[labelsep=space]{caption}
\begin{document}

\pagestyle{fancy}

\title{Insights into dendritic growth mechanisms in batteries: A combined machine learning and computational study}

\maketitle

 
\author{Zirui Zhao} 
\author{Junchao Xia} 
\author{Si Wu}
\author{Xiaoke Wang}
\author{Guanping Xu} 
\author{Yinghao Zhu} 
\author{Jing Sun} 
\author{Hai-Feng Li*} 


\begin{affiliations}
Z. R. Zhao, X. K. Wang, G. P. Xu, Dr. Y. H. Zhu, Prof. H.-F. Li \\
Institute of Applied Physics and Materials Engineering, University of Macau, Avenida da Universidade, Taipa, Macao SAR 999078, China \\
Email Address: haifengli@um.edu.mo

Dr. J. C. Xia \\
School of Materials Science and Engineering, Chang’an University, Xi’an 710061, China \\

Dr. S. Wu \\
School of Physical Science and Technology, Ningbo University, Ningbo 315211, China \\

Prof. J. Sun \\
Faculty of Data Science, City University of Macau, Avenida Padre Tomas Pereira, Taipa, Macao SAR 999078, China \\

\end{affiliations}


\begin{abstract}
In recent years, researchers have increasingly sought batteries as an efficient and cost-effective solution for energy storage and supply, owing to their high energy density, low cost, and environmental resilience. However, the issue of dendrite growth has emerged as a significant obstacle in battery development. Excessive dendrite growth during charging and discharging processes can lead to battery short-circuiting, degradation of electrochemical performance, reduced cycle life, and abnormal exothermic events. Consequently, understanding the dendrite growth process has become a key challenge for researchers. In this study, we investigated dendrite growth mechanisms in batteries using a combined machine learning approach, specifically a two-dimensional artificial convolutional neural network (CNN) model, along with computational methods. We developed two distinct computer models to predict dendrite growth in batteries. The CNN-1 model employs standard convolutional neural network techniques for dendritic growth prediction, while CNN-2 integrates additional physical parameters to enhance model robustness. Our results demonstrate that CNN-2 significantly enhances prediction accuracy, offering deeper insights into the impact of physical factors on dendritic growth. This improved model effectively captures the dynamic nature of dendrite formation, exhibiting high accuracy and sensitivity. These findings contribute to the advancement of safer and more reliable energy storage systems.
\end{abstract}


\section{INTRODUCTION}

\begin{figure*} [t]
\centering \includegraphics[width=0.82\textwidth]{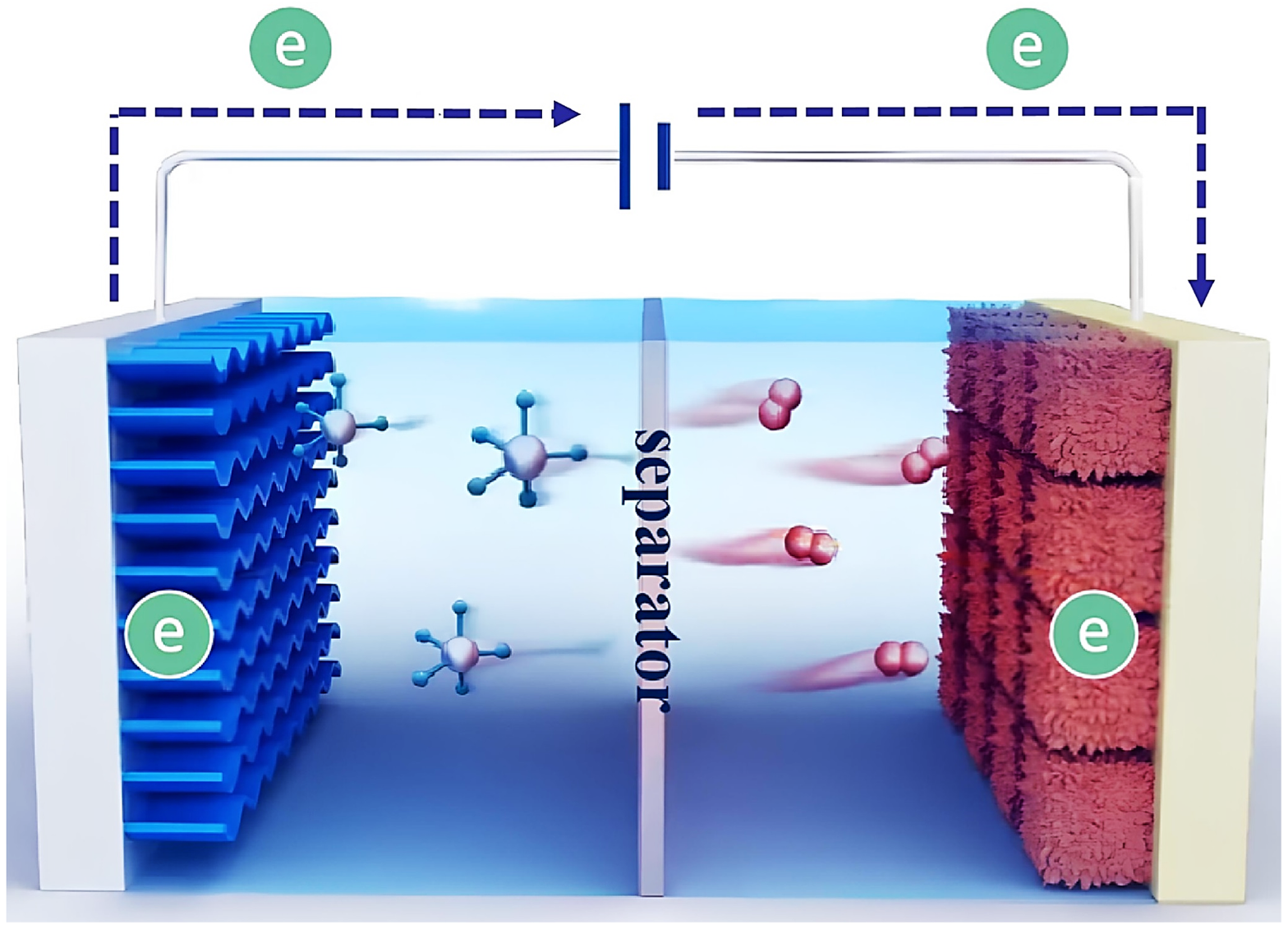}
\caption{Structure of a typical aqueous metal-ion battery.}
\label{battery}
\end{figure*}

Batteries have garnered significant interest from researchers for their exceptional energy density, portability, and safe energy supply process, making them a focal point in the development of next-generation energy storage technologies. Among these, aqueous batteries are recognized by researchers as a promising, cost-effective, and efficient energy storage and supply solution. Their low cost, high energy density, and environmental adaptability make them well-suited for various applications, including renewable energy storage and portable electronic devices.\cite{xu2024advances, chen2024metal, zhang2024smart, LIAO2022139730, ZHOU2023101248, zhou2023machine, XIAO2024151111, WANG202479, ZHAO2024112982, ZHAO202401} As illustrated in Figure~\ref{battery}, during the discharging process, metal materials at the negative electrode undergoes oxidation to generate metal ions, releasing them into the electrolyte where they migrate towards the positive electrode. Simultaneously, electrons flow through the external circuit towards the positive electrode. Conversely, during the charging process, metal ions travel from the positive electrode through the electrolyte to the surface of the metal monomer negative electrode, where they are deposited. Throughout the charging and discharging cycles, the electrolyte facilitates smooth redox reactions, providing the necessary ions for transport and enabling the battery to undergo repeated charging and discharging processes.\cite{lei2022opportunities, zou2022emerging} Therefore, the operational principle of a metal-ion battery relies on the electrolyte-mediated migration of metal ions between the positive and negative electrodes, coupled with electrochemical reactions.

\begin{figure*} [t]
\centering \includegraphics[width=0.82\textwidth]{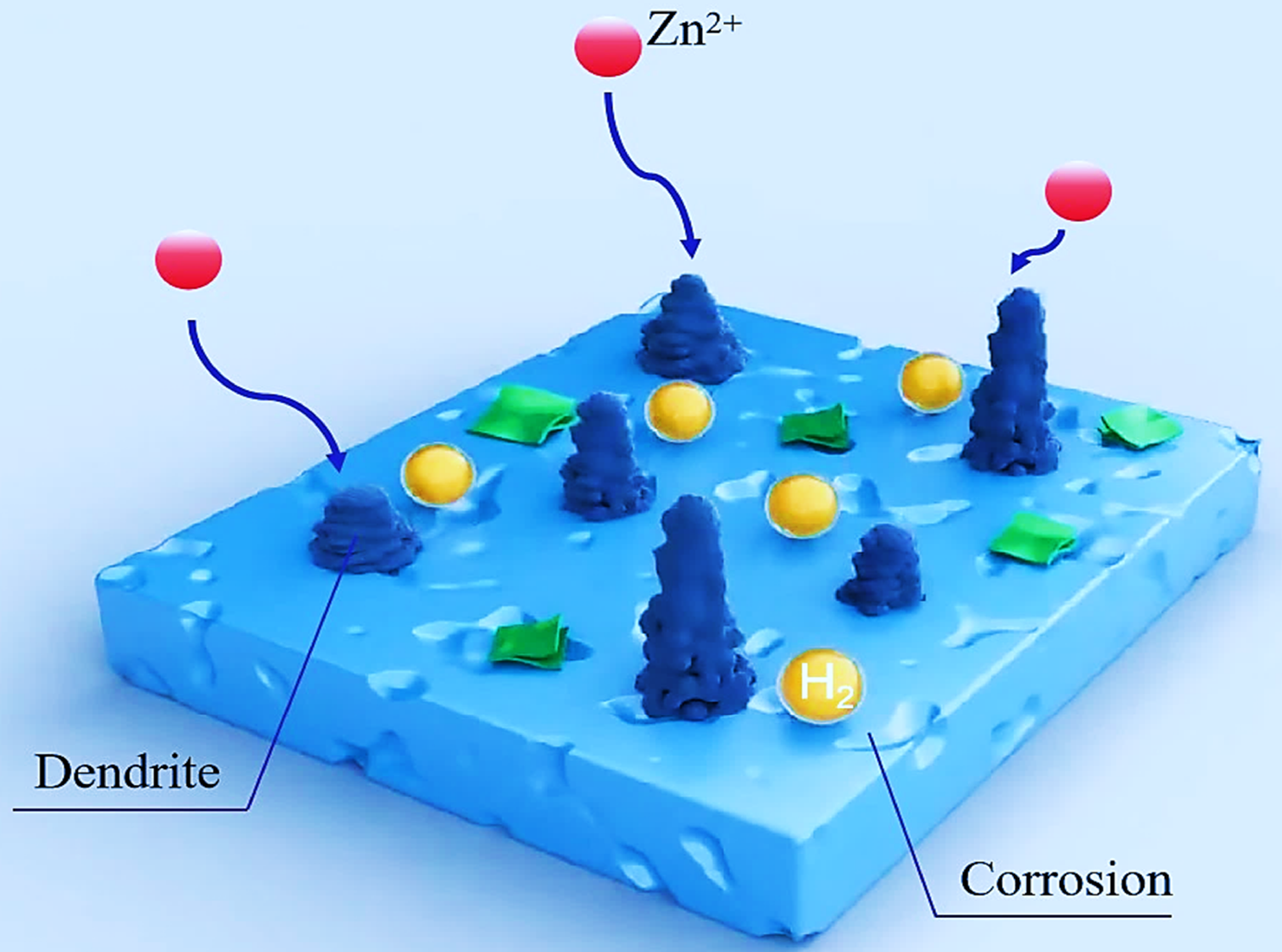}
\caption{Dendritic deposition/growth on the anode surface, utilizing aqueous metal-ion batteries as a case study.}
\label{dendritic}
\end{figure*}

The dendrite growth phenomenon plays a crucial role in battery systems.\cite{park2024fortifying, kim2024gold, song2024blocking, https://doi.org/10.1002/cey2.525, https://doi.org/10.1002/cey2.155, FANG2024143661, morey2022interfacial} During the charging and discharging processes, polarization and depolarization phenomena between the electrodes facilitate the oxidation of metal ions to metal materials. Subsequently, metal monomers are deposited in a dendrite-like structure on the surface of the negative electrode plate, leading to dendrite formation (as depicted in Figure~\ref{dendritic}). Variations in local current density and metal ion concentration during actual discharge processes increase the likelihood of metal deposition at sites with local defects or non-flat surfaces, promoting the growth of dendritic crystals. These dendrites can elongate over successive charging and discharging cycles, potentially penetrating the separator and causing short circuits, thereby compromising battery performance, reducing cycle life, and posing safety risks. Therefore, understanding and controlling dendrite growth is paramount for enhancing battery performance. Traditional analytical models for dendrite growth in batteries often deviate significantly from real experimental conditions. Early models, based on theoretical physics calculations of interfaces and local equilibrium, fail to accurately simulate and describe dendrite growth. Subsequent advancements by researchers have proposed more sophisticated frameworks, yet accuracy remains a challenge.\cite{barai2017lithium} The complexity arises from the diverse conditions governing the solid-state properties of battery materials. Factors such as interfacial dynamics, morphological stability, and solute partitioning influence ion motion rates, solidification rates, microstructure, shape retention, defect formation, and mechanical properties, posing challenges in modeling dendrite growth phenomena. Current computational models, including density functional theory, are constrained by assumptions like rotating paraboloids at solid-liquid interfaces, limiting their ability to capture the intricate dynamics of solidification. To address these limitations, researchers are exploring advanced models to enhance accuracy and predictive capabilities in describing dendrite growth phenomena.\cite{jana2019electrochemomechanics, morey2024framework} In the realm of dendrite growth evaluation, researchers have developed various numerical simulation methods such as molecular dynamics (MD), phase-field (PF), Monte Carlo (MC), cellular automata (CA), and others.\cite{li2015synergetic} For instance, the frontier tracking method minimizes total system energy while maintaining zero entropy function variability but faces challenges in computational complexity and efficiency for large-scale simulations. In contrast, MC and CA methods offer improved robustness and efficiency in predicting microstructure formation. PF methods are renowned for elucidating phase-transition physics, prompting efforts to integrate PF insights into CA and MC methods for enhanced modeling. Additionally, researchers are exploring the use of MD to describe dendrite growth in batteries, leveraging its effectiveness in simulating molecular movement at interfaces for stacking processes.\cite{crowther2008effect} However, constructing an accurate potential function to describe the entire physical system poses a significant challenge. While numerical methods can simulate complex interfacial morphologies and molecular trajectories, they are computationally intensive and may not always yield precise predictions, limiting their suitability for high-throughput calculations.\cite{liu2017model} Researchers have also endeavored to develop multiscale models by integrating numerical methods at different scales, such as finite difference, lattice Boltzmann, and fast Fourier transform, to bridge microstructural features with macroscopic properties. These models aim to capture complex material behaviors across scales but encounter challenges related to computational cost and prediction accuracy. Further studies are required to enhance the efficiency and accuracy of modeling continuous materials at macroscopic or microscopic scales to address these challenges. Aqueous batteries have attracted considerable attention as a promising solution for cost-effective and efficient energy storage, primarily due to their affordability, high-energy density, and environmental adaptability. However, these systems face significant challenges that hinder their practical application. One major limitation is the narrow thermodynamic electrochemical stability window of water, which is limited to 1.23 V. Beyond this threshold, water undergoes anodic oxidation to oxygen and cathodic reduction to hydrogen, thereby severely restricting the achievable energy density of aqueous batteries.\cite{Yu2023, D3CP02162A} Additionally, aqueous batteries experience slow reaction kinetics and structural instability, further complicating performance optimization.\cite{zhang2024rational, yu2024theoretical} Overcoming these limitations is crucial for advancing the development of safer, higher-performance aqueous battery technologies.

The rapid advancement of machine learning technology in recent times, coupled with the increasing complexity of these models, has led to the emergence of more efficient data-driven methods compared to traditional empirically-driven, theoretically-driven, and computationally-driven approaches. Researchers have leveraged these advancements to catalyze a significant paradigm shift in materials research.\cite{wang2018stress} Machine learning (ML) techniques, fundamentally data-driven modeling methods, utilize data to construct surrogate models that predict target properties based on material descriptors. ML techniques excel in handling high-dimensional mathematical regression problems and offer significantly faster computational speeds than traditional theoretical or numerical methods, positioning them as pivotal tools in materials research. These techniques have been widely adopted in materials research, yielding significant results across various application scenarios. However, ML-driven models face challenges such as small datasets and substantial experimental errors,\cite{he2021understanding} which can impede the effectiveness and reliability of ML models in materials research. Researchers are actively exploring methods and strategies to overcome these limitations.\cite{brissot1999dendritic} To enhance the reliability and accuracy of ML models in materials research, researchers are focusing on improving the predictive performance of data-driven machine learning (DDML) methods. Techniques such as data augmentation and feature engineering are commonly employed for this purpose, involving methods like artificially expanding dataset sizes by generating modified data points or optimizing feature selection, extraction, and transformation to enhance model performance. Additionally, the introduction of transfer learning has provided researchers with a mechanism to transfer knowledge gained from one ML task to improve the performance of related tasks.\cite{wang2012thermal, cannon2023smoothed} Despite the benefits of transfer learning, ML models are often perceived as black boxes, lacking transparency in their operations and extrapolation capabilities. Consequently, researchers, including those in this study, are exploring avenues to address this issue by integrating fundamental physical principles into the learning process. By infusing models with the ability to harness the rich information embedded in physics-based models, this approach, known as Physics-Informed Machine Learning (PIML), not only enhances predictive capabilities but also offers high model interpretability. The integration of PIML combines the flexibility and adaptability of a data-driven approach with the rigor of physics-based theoretical calculations, enabling the resolution of complex problems and advancing the frontiers of materials science research \cite{crowther2008effect}. Recent advances in machine learning have significantly impacted the field of energy storage, especially concerning aqueous batteries. For instance, solubility prediction for energy storage materials has been effectively implemented using deep learning techniques, providing valuable insights into optimizing battery components.\cite{chaka2023advancing} Furthermore, density functional theory (DFT) calculations and thermodynamic integration methods have been employed to study electron and ion transfer processes. These methods offer a deeper understanding of the interactions between lithium metal and electrolytes in battery systems.\cite{angarita2020insights} Such studies have informed and enriched the development of our machine learning model, enabling us to incorporate similar approaches to enhance the predictive accuracy of dendrite growth in aqueous batteries.

In the current research landscape, ML-based models focusing on dendrite growth phenomena have emerged in recent literature.\cite{ely2013heterogeneous, pei2017nanoscale, shen2019magnetic} These data-driven models, developed by researchers, span a broad spectrum of scales from microscopic to macroscopic and often incorporate multiphysics field methods. Their objective is to forecast dendrite growth phenomena under varying physical conditions, such as the growth rate of undercooled multicomponent metal alloys, the occurrence of pore defects during the solidification of aluminum alloys, and the susceptibility to solidification cracking.\cite{ely2013heterogeneous} However, recent studies have identified significant limitations in these models, as many rely solely on a data-driven approach without adequately considering the influence of physics on the phenomenon. This oversight particularly impacts the accurate prediction of solid-liquid interfacial velocities, especially in scenarios involving dendritic growth in binary alloying systems lacking explicit alloying constraints. 
Dendritic growth in the field of battery research remains a critical issue impacting battery performance and safety. Despite extensive efforts to elucidate the mechanisms of dendritic growth, traditional methods reliant on experiments and theoretical calculations have inherent limitations.\cite{gupta2023estimation} Our study builds upon this foundation with innovative advancements. To enhance the accuracy of our model, we integrated machine learning with computational simulations. Our model achieved a remarkable accuracy of 0.80, representing a significant improvement of nearly 20\% over conventional methods. This heightened accuracy can be attributed to the capabilities of machine learning algorithms in big data analysis and pattern recognition, enabling precise prediction and simulation of dendritic growth processes.\cite{brissot1999dendritic} Furthermore, our study transcended data and models by incorporating visualization techniques to visually represent the dendritic growth process. Through high-resolution videos, we dynamically illustrate the entire process of dendrite formation and expansion. This visualization approach not only aids in comprehending the intricate mechanisms of dendritic growth but also serves as a valuable reference for future research and applications. Our research achieves new levels of accuracy and intuitiveness, presenting an innovative and effective approach to dendritic studies in the battery field.\cite{barai2017lithium}

This study aims to integrate data-driven machine learning with physics-based modeling to enhance our understanding and predictive capabilities in the critical field of materials science. By bridging this gap, the study seeks to develop machine learning-based models in conjunction with Vienna Ab initio Simulation Package (VASP) calculations to simulate dendrite growth phenomena in aqueous cells. Additionally, the study aims to conduct a comprehensive comparison between data-driven machine learning (DDML) and Physics-Informed Machine Learning (PIML) approaches. Among the evaluated models, Model 2 emerged as the most suitable method. This outcome underscores the potential of machine learning-based models in advancing our comprehension of dendrite growth in aqueous cells, emphasizing the significance of integrating both data-driven and physics-informed approaches in future research endeavors.

\section{EXPERIMENTAL SECTION}

The workflow for our CNN-based machine learning model is summarized as follows. First, data were collected from various sources, including experimental studies and computational simulations, focusing on key input parameters such as electrolyte concentration, current density, and temperature. These data were then preprocessed using custom Python scripts to standardize the input variables, including scaling and unit conversion. The preprocessed data were divided into training, validation, and testing sets, providing a solid foundation for subsequent model training.

The CNN model comprises eight convolutional layers and eight pooling layers. The input includes battery components and associated physical parameters, while the output predicts the dendritic growth pattern, with dendritic length at specific time intervals as the primary metric. The model was trained using backpropagation and gradient descent techniques, followed by validation and testing on unseen data. The model's predictions were compared with experimental results to ensure reliability.

\subsection{Data collection and preparation}

Acquiring a comprehensive dataset of battery dendrite studies is essential for training and validating data-driven models. In this section, we employed a web-scraping approach using Python's Selenium library to extract relevant data from publicly available websites and compile our database. In our approach, we utilized a custom-built convolutional neural network model and a theoretical modeling model based on the traditional VASP software. The artificial neural network model was trained using data collected from 6514 dendrite growth processes. The model's performance on both the test set and the training set indicates that the dataset size is adequate for model training. To gather literature on battery dendrite growth phenomena, we accessed reputable academic databases and repositories, obtaining publications such as journals, conference proceedings, and research articles.\cite{jana2017lithium, jana2019electrochemomechanics, yan2018computational, akolkar2014modeling, wang2020suppressing, xu2018recent, hong2018phase, yan2019temperature, steiger2014mechanisms, rosso2001onset, barai2018impact, ferrese2012lithium, kushima2017liquid, wood2017lithium, tan2020simulating} Given the abundance of relevant literature, we employed an automated web scraping approach using the Selenium library in Python.\cite{percival2014test} This tool facilitated systematic browsing of selected online databases, replicating a user's interactive download process to extract information from web pages and initiate the database construction.\cite{aryanfar2014dynamics} Python scripts were developed to interact with the browser, enabling dynamic and efficient data collection by browsing search results, accessing individual publication pages, and extracting metadata like titles, authors, abstracts, and publication dates to compile large-scale datasets.\cite{zhang2019dendrites, xu2017suppression} Approximately 6000 research entries were collected to form a comprehensive dataset on dendritic growth processes. The dataset size was deemed suitable for capturing the complexities of dendritic growth, crucial for training our machine learning model and enhancing its predictive capabilities. In addition, our dataset comprises experimental and simulation data from both zinc-ion (Zn-ion) and lithium-ion (Li-ion) batteries, encompassing a wide range of parameters such as operating conditions and electrolyte compositions. And to ensure compatibility across different battery chemistries, all features were normalized, and categorical variables (cation type) were encoded using one-hot encoding. This preprocessing step allows the model to effectively learn the distinct dendritic growth mechanisms associated with each battery type. Following successful data extraction, thorough pre-processing was conducted to organize, extract relevant information, and structure the data into a consistent format suitable for further analysis. Duplicate entries were eliminated, and missing data were addressed to ensure dataset completeness and quality.

Further more, data processing plays a key role in ensuring the quality and robustness of the model’s predictions. During the segmentation of the data, several preprocessing techniques were applied to improve the overall dataset quality. To address potential overfitting and ensure the model's generalization capability, data augmentation techniques were employed. These included random rotations, flips, and scaling of the input images, which simulated variations in dendritic growth and allowed the model to learn more diverse patterns. This strategy was especially important for enhancing the model’s ability to generalize when exposed to unseen data.

Moreover, we implemented outlier processing to eliminate data points that could potentially skew the model’s learning process. Outliers, often the result of measurement errors or anomalous data, were detected and excluded using robust statistical methods such as interquartile range (IQR)-based filtering. By removing these outliers, we ensured that the model was trained on data that better reflected realistic dendritic growth behavior. In addition, noise reduction techniques were applied to remove high-frequency noise present in the data. Gaussian smoothing and median filtering were particularly useful in mitigating sensor or measurement noise, allowing the model to focus more effectively on the essential features of dendrite morphology rather than irrelevant variations.

A significant component of our model was the use of a customized loss function, which was tailored to better capture the complexities of dendritic growth. While a standard loss function like mean squared error (MSE) could be used for general regression tasks, we found it inadequate for dendritic growth, which involves highly nonlinear and irregular patterns. To address this, we designed a loss function that combined MSE with a regularization term that penalized deviations in dendrite morphology. This regularization term ensured that the model did not overfit to minor noise or irrelevant features, but instead focused on the core patterns of dendritic growth, thus improving the model’s robustness.

In terms of convergence, the training process was closely monitored using both the validation loss and accuracy. Convergence was deemed to have occurred when the validation loss stabilized and showed minimal fluctuations over multiple epochs. To avoid overfitting, early stopping was implemented, which halted training when there was no significant improvement in the validation loss for a predefined number of epochs. This strategy helped ensure that the model achieved optimal performance without fitting too closely to the training data, ultimately leading to better generalization on unseen data.

By providing these details on data processing, including augmentation, outlier removal, noise reduction, and the customized loss function, we aim to offer a clearer understanding of the steps taken to optimize the model and improve its accuracy in predicting dendritic growth.

In this study, we refer to 6514 dendrite growth processes as a comprehensive database compiled from various published studies and articles by different researchers. This database consists of images capturing dendrite growth at different time points under various experimental conditions. The images were collected from existing literature to build a robust dataset for our modeling approach. This extensive dataset allows us to create a versatile and accurate model by leveraging the diverse experimental conditions and outcomes reported by multiple research groups.

\subsection{Architecture of convolutional neural networks}

To enhance the accuracy of predictions and advance our comprehension of cell dendrite growth processes, we employed convolutional neural networks (CNNs).\cite{girshick2015fast, he2017mask} This section offers an in-depth exploration of the CNN architecture implemented in our study, detailing its design, training methodology, and significance in predicting cell dendrite growth phenomena. We acknowledge the concern regarding the clarity of the machine learning model description. The model we used for predicting dendritic growth is an artificial convolutional neural network (CNN), as shown in the schematic diagram. This model features an architecture with eight convolutional layers and eight pooling layers. We acknowledge the lack of detail in the initial version of the manuscript and will include a more comprehensive description of the model architecture and its relevance to dendritic growth prediction in the revised version.

\textbf{Architecture:} Our CNN model is built upon the UNet architecture, a convolutional neural network renowned for its efficiency and accuracy in handling large-scale two-dimensional matrix-like data. The distinctive U-type architecture of UNet enables it to capture intricate features at both local and global scales, making it well-suited for the structure and composition of battery dendrite growth datasets.

\textbf{Input and output:} 
The crystal structure parameters utilized in our study inherently possess three-dimensional characteristics, capturing the intricate details essential for accurately modeling dendritic growth in batteries. However, working directly with three-dimensional data presents computational challenges. To address this, we employ a method where multiple two-dimensional crystal structure parameters are overlaid to construct three-dimensional structures. 

\textbf{Superimposition of two-dimensional parameters:}
Each set of two-dimensional parameters represents a cross-section of the overall crystal structure. Through precise alignment and integration of these cross-sections, we reconstruct the complete three-dimensional crystal structure. This layered approach ensures the accurate representation of spatial relationships and interactions within the crystal structures. The process involves two key steps: Data Alignment and Integration. During Data Alignment, each two-dimensional slice is aligned based on specific reference points within the crystal structure to maintain consistency. Subsequently, the aligned slices are superimposed to generate the three-dimensional structure, preserving the integrity and spatial accuracy of the data. The process of constructing 3D crystal structures from 2D parameters involves several key steps. First, the 2D parameters, such as lattice dimensions and symmetry, are obtained from computational models or experimental data. For example, in the case of lithium-ion battery anodes, the 2D parameters could include the hexagonal arrangement of lithium ions in the layered structure of graphite, as well as the interlayer distance between carbon atoms. These 2D parameters form the basis for constructing a 3D model of the anode material. To create a three-dimensional structure, these 2D parameters are extended by applying spatial transformations that replicate the crystal’s symmetry and atomic interactions. In lithium-ion battery materials like graphite, this involves stacking 2D graphite layers along the z-axis while maintaining the correct interlayer spacing and accounting for van der Waals interactions between the layers. By incorporating these 2D parameters, we can accurately construct a 3D model that reflects the material's atomic configuration, providing a detailed representation of how lithium ions move within the anode during charging and discharging processes.

The CNN receives a multidimensional array as input, representing the crystal structure at various time steps. Its output is intended to forecast the future state of cell dendrite growth, thereby effectively capturing the evolving morphology over time. And what needs to be raised is that, in this study our primary focus is on the dendrite growth phenomenon, which exhibits similar fundamental characteristics across different types of batteries, including both zinc-ion and lithium-ion batteries. Our research employs a comprehensive modeling approach that is not strictly limited to a specific type of battery. Instead, our model can be adjusted according to different parameters to accommodate various battery systems.

This flexibility is crucial for our study as it allows us to achieve a broader and more generalized understanding of dendrite growth mechanisms. When collecting and establishing our database, we included data from multiple battery types to ensure the wide applicability of our model. The main difference between zinc-ion and lithium-ion batteries lies in the type of electrolyte used—aqueous electrolytes for the former and non-aqueous electrolytes for the latter. However, these differences are reflected in our model as parameter adjustments rather than fundamental methodological changes.

Therefore, while we references literature related to both zinc-ion and lithium-ion batteries in different sections, this does not affect the applicability and consistency of our model. To ensure coherence and clarity in our narrative, we unified the relevant terminology and discussions throughout this study. Additionally, we acknowledge that the current study primarily outputs results for lithium-ion batteries. This choice was made to highlight specific aspects of our findings.

\textbf{Convolutional layers:} The primary neural network architecture comprises a sequence of convolutional layers (13 in total) that utilize filters to analyze and extract features from the input data. This process enables the model to acquire a hierarchical representation of the crystal structure.

\textbf{Downsampling and upsampling:} The UNet architecture employed in this study comprises two paths: downsampling and upsampling, realized through four pooling layers and four transposed convolutional layers. Downsampling via pooling layers captures high-level features, while upsampling through transposed convolutional layers reconstructs the output at an increased spatial resolution.

\textbf{Loss function and optimization:} To guide the training process, a customized loss function tailored to the intricacies of cell dendrite growth prediction is utilized. Achieving a balance between rapid convergence and stability, the percentage of similarity to experimental growth serves as the output data, directing the training process and optimizing it through stochastic gradient descent.\cite{barron2019general}

\textbf{Training and validation:} The CNN model underwent comprehensive training using a dataset derived from literature and a simulated dataset of cell dendrite growth. Validation was conducted on unseen data to evaluate the prediction accuracy of the model across various conditions.\cite{barron2019general}

The dataset was partitioned into training and validation sets to facilitate model training and assessment. Specifically, an 80/20 split was employed, allocating 80\% of the data (5212 samples) for model training and reserving the remaining 20\% (1302 samples) for validation purposes. During the hyper-parameter tuning phase, the training set was further subdivided into fractional training sets to enable cross-validation and optimize the model's performance. The training process encompassed multiple iterations of hyper-parameter tuning and cross-validation, leveraging advanced machine learning techniques such as grid search and k-fold cross-validation (k = 5) to systematically assess and select the most effective model configurations. Various metrics, including accuracy, precision, recall, and F1-score, were utilized to evaluate the model's performance. The final model achieved an accuracy of 0.80 on the validation set, signifying a notable enhancement over conventional methods. To reinforce our conclusions and enhance interpretability, we generated high-resolution videos illustrating the dendritic growth processes. These videos, produced using the trained model, offer a visual depiction of the dynamic characteristics of dendrites.

Overall, our CNN model is based on the UNet architecture and, following training, demonstrates the capability to predict the cell dendrite growth process effectively, thereby achieving the objective of enhancing crystal growth prediction accuracy.\cite{guan2019fully}

\subsection{Model evaluation and validation}

Evaluating CNN models for cell dendrite growth prediction necessitates a thorough examination of their robustness and accuracy across diverse datasets. Techniques such as sensitivity analysis and cross-validation with varied datasets are employed to validate the model's reliability and suitability for real-world applications. This section focuses on the methodologies and metrics utilized to assess the performance of the cell dendrite growth prediction model.

To evaluate the reliability of the proposed model, we used the dendrite length as the primary characteristic parameter to describe dendritic growth. This parameter was measured at 30-second intervals during the battery's operation, providing a consistent and quantifiable metric for comparing the model's predictions with experimental data. By analyzing dendritic growth over time, we assessed the model's accuracy. The results showed a strong correlation between the predicted dendrite lengths and the experimental observations, demonstrating the model's robustness in capturing the key dynamics of dendritic growth.

\textbf{Training and testing split:} A rigorous evaluation process was implemented to gauge the robustness of our CNN model in predicting dendritic growth. The dataset was partitioned into training and testing sets to assess the model's generalization capacity. This partitioning ensured the CNN model's proficiency in providing accurate predictions for new samples.

\textbf{Evaluation metrics:} 
In our analysis, we evaluated the model performance using the \emph{R}-squared (\( R^2 \)) and Mean Squared Error (MSE) metrics. The formulas for these metrics are as follows:

\[
R^2 = 1 - \frac{\sum_{i=1}^{n} (y_i - \hat{y}_i)^2}{\sum_{i=1}^{n} (y_i - \bar{y})^2},
\]
where \( y_i \) are the observed values, \( \hat{y}_i \) are the predicted values, \( \bar{y} \) is the mean of the observed values, and \( n \) is the number of observations.

\[
\text{MSE} = \frac{1}{n} \sum_{i=1}^{n} (y_i - \hat{y}_i)^2,
\]
where \( y_i \) are the observed values, \( \hat{y}_i \) are the predicted values, and \( n \) is the number of observations. Incorporating these formulas offers a precise insight into the statistical metrics utilized to assess the accuracy and reliability of our machine learning model in predicting dendritic growth. These metrics offer a thorough evaluation of the model's accuracy, precision, and overall performance in predicting dendritic growth patterns.

\textbf{Cross-validation:} To validate the model and mitigate overfitting, we implemented k-fold cross-validation. This method partitions the dataset into k subsets, training the model on k-1 subsets and validating it on the remaining subset. This iterative process was repeated k times to ensure consistent performance across different data subsets.\cite{browne2000cross}

\textbf{Comparison with baseline models:} The performance of the CNN model was compared against traditional simulation-based methods and simpler machine learning approaches.\cite{tan2020simulating} This comparative analysis enabled the assessment of the CNN model's efficacy and superiority in predicting dendritic growth compared to established methods.

\textbf{Sensitivity analysis:} A sensitivity analysis was conducted to assess the model's response to variations in input parameters. This analysis provided insights into the model's robustness and its capability to accurately predict dendritic growth under diverse conditions.

\textbf{Real-world applicability:} To evaluate the CNN model's real-world utility, we juxtaposed its predictions with experimental observations. This comparison validated the model's accuracy in predicting dendritic growth patterns, affirming its reliability and adaptability in practical scenarios.

Our CNN model underwent a comprehensive evaluation, incorporating a range of metrics and validation techniques. This rigorous assessment ensured the reliability and adaptability of the CNN model in predicting dendritic growth patterns, establishing it as a valuable tool for comprehending and forecasting dendritic growth across various contexts.

\section{RESULTS AND DISCUSSION}

\subsection{VASP simulation}

In conjunction with the literature dataset, we enriched our study by generating data on cell dendrite growth using VASP software. This section delineates the methodology employed to construct this synthetic dataset, which served as a valuable complement to our empirical data.

The establishment of a VASP simulation necessitates the specification of a series of parameters to govern the conditions of dendritic growth. These parameters encompass temperature, pressure, substrate properties, and initial crystal configuration. Leveraging the advanced atomic number simulation software package VASP, we elucidated the electronic structure of the crystal system, capturing the intricate interactions that dictate the growth of cell dendrites. This computational model enables the simulation of the crystal structure's evolution over time. The simulation progresses through iterative time steps, with each step representing a discrete moment in the cell dendrite growth process. At each time step, VASP computes the forces and energies exerted on the atoms, directing their movement and influencing the overall crystal structure. The VASP simulation yields a substantial volume of data detailing the progression of the cell's dendrite growth process, providing insights into the positions, energies, and structures of the atoms as they evolve over time.

The VASP simulations in this work, while computationally demanding, were essential for capturing atomic-scale mechanisms of dendritic growth. To enable direct comparison with experimental data and machine learning predictions, we focused on simplified systems, such as small-scale Li-metal/electrolyte interfaces and early-stage dendritic nucleation sites. These systems were chosen to balance computational feasibility with the ability to observe clear dendritic phenomena.

For example, in our simulations of Li-metal anodes, we observed the formation of dendritic protrusions within 10-15 ionic relaxation steps, each requiring approximately 100 CPU hours. This duration was sufficient to capture the initial stages of dendrite formation, which are critical for understanding the underlying mechanisms.

To optimize resource allocation, we employed a hierarchical approach. First, we set the initial screening to determine metastable conformations using classical molecular dynamics (MD) simulations. We then refine the DFT in a targeted manner, performing high-precision DFT calculations only for critical conformations (those near dendritic nucleation sites).

Integrating the data from simulated cell dendrite growth with the literature-based dataset establishes a foundation for a comprehensive analysis that leverages both real-world observations and the controlled environment afforded by simulation. To ensure the authenticity and reliability of the simulation data, we validate them against experimental observations from the literature-based dataset.

Modeling battery dendrite growth through this amalgamation of literature-based data and the utilization of VASP not only expands the scope of our dataset but also enhances our understanding of cell dendrite growth phenomena in greater detail. Furthermore, it augments the predictive capacity of the model across diverse conditions and scenarios.

\subsection{Machine learning modes}

\begin{figure*} [t]
\centering \includegraphics[width=0.82\textwidth]{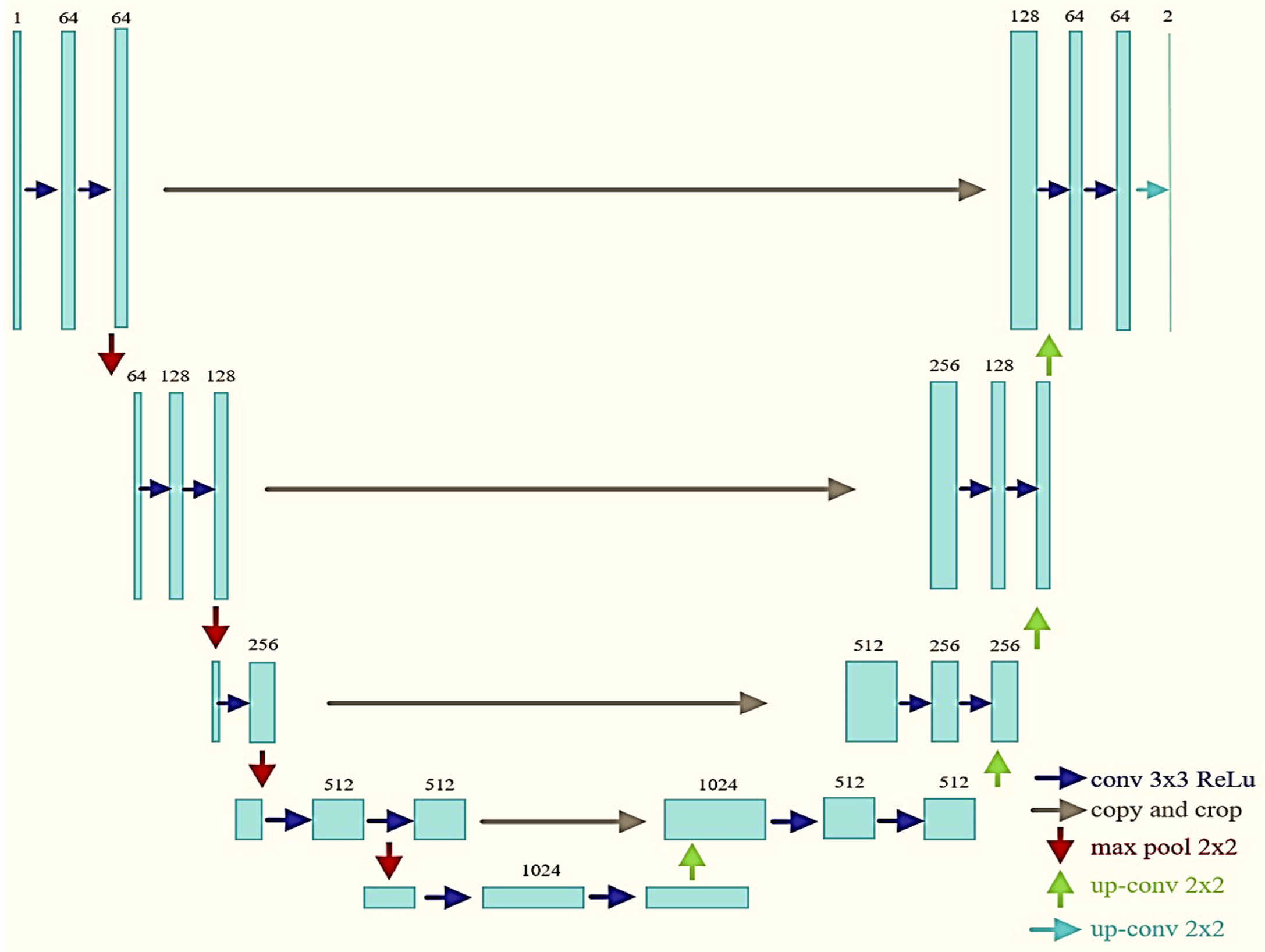}
\caption{2D convolutional neural network constructed for predicting dendritic growth processes based on simulated data collected.}
\label{2D}
\end{figure*}

\begin{figure*} [t]
\centering \includegraphics[width=0.82\textwidth]{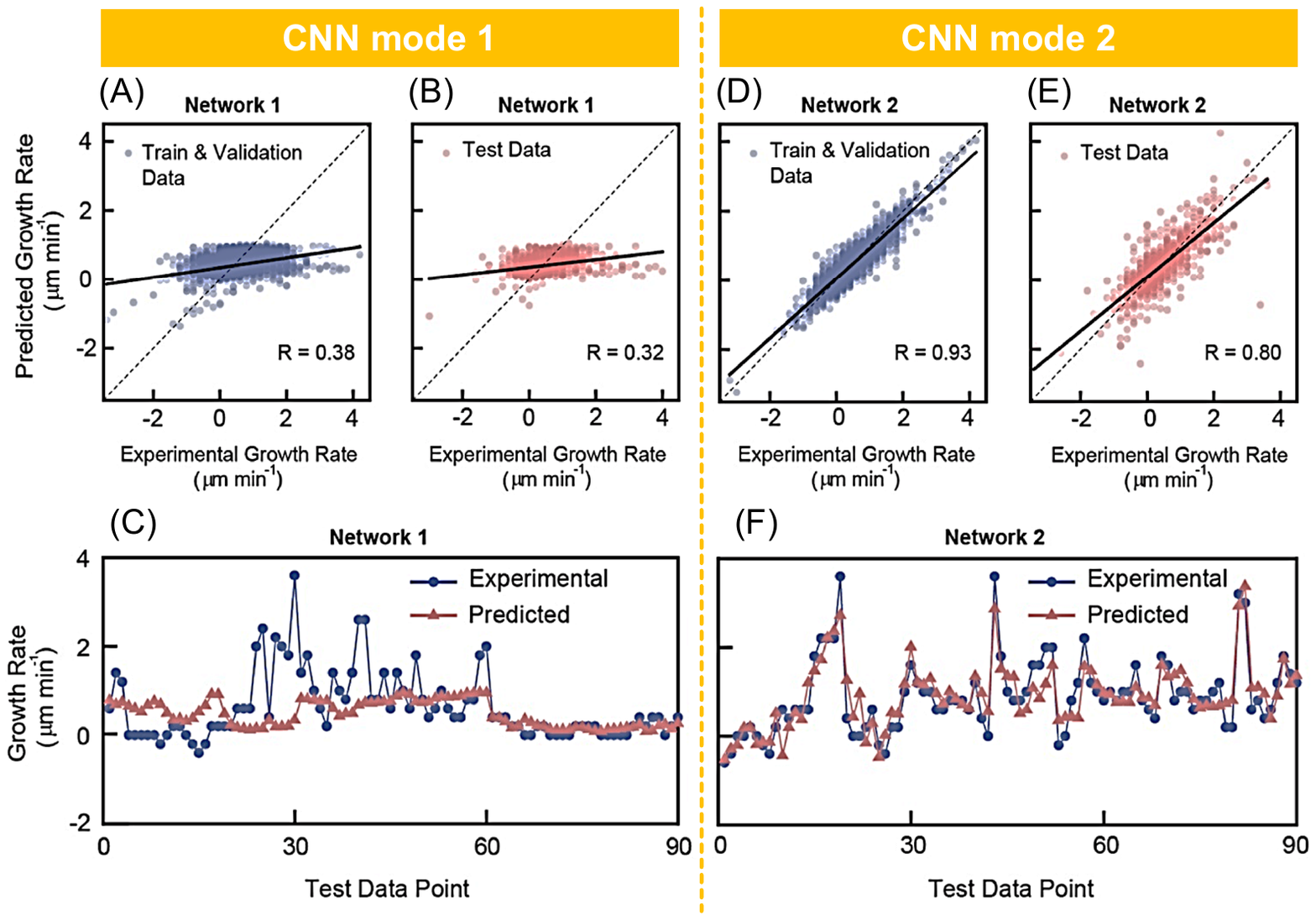}
\caption{Performance evaluation of the dendritic growth modes utilizing our trained artificial convolutional neural network. (A) and (B) illustrate the dendritic growth rates of network 1 (CNN mode 1) (omitting various physical variables), alongside their respective prediction accuracies in both training (A) and testing (B) datasets. (C) Specific performance metrics for the predicted data points. (D) and (E) The dendritic growth rates of network 2 (CNN model 2) (considering specific physical variables as input), along with their corresponding prediction accuracies in the training (D) and testing (E) datasets, respectively. (F) Additional demonstration of the performance on the predicted data points. The CNN model-2 architecture integrates physical quantities. This neural network model incorporates seven selected physical parameters as inputs, facilitating a more thorough analysis of their influence on dendritic growth rates. The incorporation of these parameters is intended to bolster the predictive capability of the CNN model by integrating pertinent physical insights.
}
\label{CNNmodes12}
\end{figure*}

In this study, a two-dimensional artificial convolutional neural network model was employed to predict the specific process of dendrite growth on battery anodes. The architecture of this network is illustrated in Figure~\ref{2D}. Two distinct networks were constructed as depicted in Figure~\ref{CNNmodes12}: network 1 (representing dendrite growth pattern 1) and network 2 (representing dendrite growth pattern 2). The accuracy metric (\emph{R}-value) for dendritic growth pattern 1 extracted through network 1 (without considering various physical variables) is notably below 1, with \emph{R} = 0.38 for the training dataset (Figure~\ref{CNNmodes12}(A)) and \emph{R} = 0.32 for the test dataset (Figure~\ref{CNNmodes12}(B)). Furthermore, a substantial disparity between the experimental and predicted data points is observed (Figure~\ref{CNNmodes12}(C)), indicating that Network 1 may not be suitable for studying dendrite growth. Conversely, the CNN model trained by incorporating specific physical variables in Network 2 demonstrates high accuracy, achieving 0.93 for the training set (Figure~\ref{CNNmodes12}(D)) and 0.80 for the test set (Figure~\ref{CNNmodes12}(E)). This suggests excellent agreement between the experimental and test data points (Figure~\ref{CNNmodes12}(F)), affirming the successful training of a CNN model using Network 2. Additionally, beyond achieving commendable accuracy metrics, our CNN model underwent a comprehensive sensitivity analysis to assess its robustness across varying input conditions.

In contrast to CNN-1, CNN-2 integrates the impact of physical variables on dendritic growth phenomena. By incorporating seven carefully selected physical parameters as inputs, CNN-2 is designed to more accurately capture the fundamental physical processes. 

The selection of the seven physical parameters used in the CNN-2 model—temperature, concentration gradient, electric field strength, ion concentration, surface energy of dendrites, fluidity of the solution, and the number of lattice defects—was based on their critical influence on dendritic growth and the electrochemical dynamics within the system. Each of these parameters directly impacts the rate, pattern, and stability of dendrite formation. Temperature is a fundamental factor that affects ion mobility and electrochemical reactions, thus influencing dendrite growth kinetics. The concentration gradient drives the redistribution of ions, contributing to uneven deposition and potential dendrite formation. Electric field strength plays a pivotal role in the electrochemical driving force for ion movement, directly impacting the growth direction and morphology of dendrites. Ion concentration determines the availability of ions for deposition, which is essential for dendritic structure formation. The surface energy of dendrites governs the stability of their growth; high surface energy can lead to unstable dendritic structures, while lower surface energy may promote more uniform growth. The fluidity of the solution affects the transport of ions and can influence the dissolution or growth of dendrites, depending on the solution's viscosity. Finally, the number of lattice defects is important for nucleation, as defects in the lattice can act as preferential sites for dendrite formation, promoting their growth. Together, these parameters provide a comprehensive representation of the factors influencing dendritic growth and were selected to capture the key physical and electrochemical behaviors of the system under investigation.
This integration enhances the model's resilience, enabling CNN-2 to offer more reliable and precise forecasts of dendritic growth rates. The inclusion of these parameters helps unravel the complex interactions governing dendritic formation, making CNN-2 a more powerful tool for analyzing and predicting dendritic growth in battery systems.

Our study introduces several innovative aspects in the CNN-2 model. While previous studies have explored data-driven methodologies, we are pioneers in employing a CNN model that integrates physical parameters as inputs. Unlike conventional purely data-driven models and solely theoretical computational models,\cite{yan2018computational, akolkar2014modeling} our model demonstrates improved robustness, providing a more authentic representation of dendritic growth phenomena. These findings underscore the potential of machine learning models in enhancing our understanding and prediction of battery behavior, which is crucial for developing more efficient and reliable energy storage systems.

To provide a broader context for the contributions of this study, we further compare CNN-2 with traditional simulation-based methods, as well as other machine learning models. While CNN-2 shows a clear improvement in accuracy and generalization over CNN-1, it is essential to evaluate its performance in comparison with other established methods.
Traditional simulation-based techniques, such as Molecular Dynamics (MD), Density Functional Theory (DFT), Phase Field (PF), and Monte Carlo (MC), have long been used to model dendritic growth. These methods are known for their precision, especially at the atomic level, and are highly valuable for gaining insights into the fundamental mechanisms of dendrite formation. However, they are computationally expensive and time-consuming, especially when applied to large-scale systems or real-time predictions, making them less suitable for high-throughput applications, such as battery optimization.
In contrast, machine learning models, including Convolutional Neural Networks (CNN), offer substantial advantages in terms of computational efficiency and scalability. Specifically, CNN-2, with its more complex architecture, not only provides a more accurate representation of dendritic growth patterns compared to CNN-1, but also delivers predictions much faster than traditional methods. This enables real-time applications and faster iteration cycles in battery development, where large datasets and the need for rapid decision-making are common.
To further illustrate the advantages of CNN-2, we conducted a comparison with other popular machine learning models, such as Random Forest (RF) and Support Vector Machine (SVM). Although RF and SVM models perform well for many classification tasks, they are less effective than CNN-2 in capturing the intricate, spatially distributed nature of dendritic growth. CNN-2's deep learning framework allows it to learn complex, nonlinear relationships between physical parameters and growth patterns, making it superior in this particular context.
By presenting this comparative analysis, we aim to highlight not only the superiority of CNN-2 over CNN-1 but also its advantages over traditional simulation techniques and other machine learning models in predicting dendritic growth. This further underscores the potential of CNN-2 as an efficient and powerful tool for modeling and optimizing battery performance in real-world applications.

\begin{figure*} [t]
\centering \includegraphics[width=0.82\textwidth]{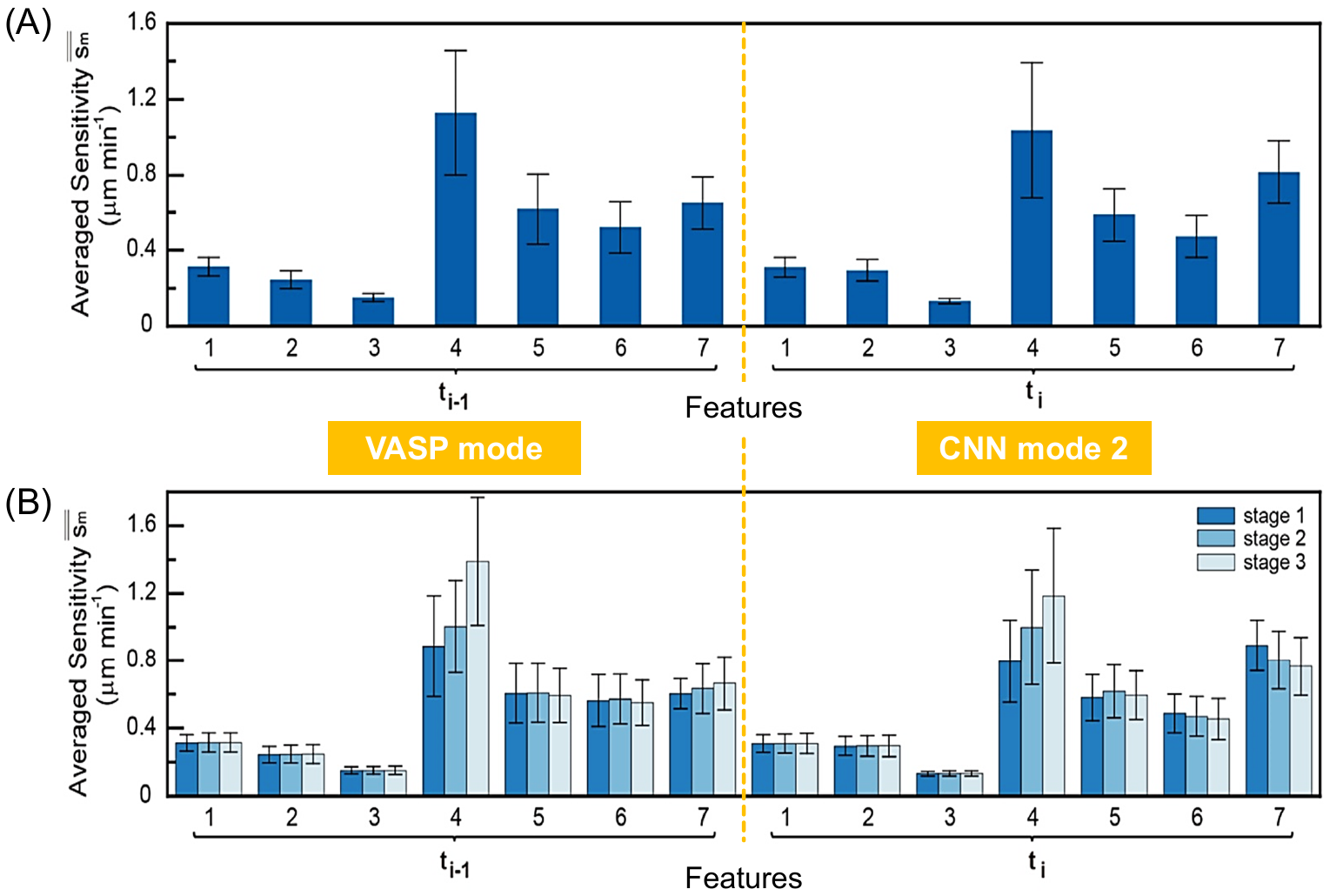}
\caption{Sensitivity assessment of the two modes is highlighted. (A) The overall assessment. (B) The sensitivity assessment across trickle charging (stage 1), constant current charging (stage 2), and constant voltage charging (stage 3). The left panels in (A) and (B) represent the assessments using VASP mode without data-driven computation, whereas the right panels correspond to CNN mode 2 with data-driven computation. The evaluation incorporates seven distinct physical parameters (t$_{\textrm{i}-1}$: VASP mode; t$_\textrm{i}$: CNN mode 2). The specific parameters are as follows, 1: temperature, 2: concentration gradient, 3: electric field strength, 4: ion concentration, 5: surface energy of dendrites, 6: fluidity of the solution, and 7: number of lattice defects.}
\label{VASPmode}
\end{figure*}

In addition, the VASP software was utilized to investigate dendrite growth dynamics. Figure~\ref{VASPmode} illustrates a thorough sensitivity assessment (Figure~\ref{VASPmode}(A)) and an in-depth evaluation of three stages (Figure~\ref{VASPmode}(B)): trickle charging (stage 1), constant current charging (stage 2), and constant voltage charging (stage 3). 

These steps are three key stages of dendrite growth during the charging process. Trickle charging is a low-rate, slow charging method typically used to maintain the battery’s charge without overcharging, which helps mitigate dendrite formation in the early stages. In contrast, constant current charging maintains a steady current throughout the charging process, but can lead to dendrite growth if the metal ion deposition becomes uneven. Finally, constant voltage charging keeps the voltage constant while the current gradually decreases as the battery reaches full charge. This stage is particularly sensitive to dendrite growth, as any voltage instability can result in excessive deposition of metal, promoting dendrite formation. Understanding these charging stages is crucial for controlling dendritic growth and improving battery performance and safety.

We mentioned that CNN-2 integrates additional physical parameters to enhance its robustness. To clarify and provide more detail, we include specific examples of these parameters. For instance, ion concentration and surface energy of dendrites are two crucial parameters incorporated into CNN-2. Ion concentration directly affects the electrochemical gradient, influencing ion deposition rates and dendrite growth behavior. Surface energy of dendrites plays a critical role in determining the morphology and stability of dendrites, as it governs the interaction between dendrite tips and the surrounding electrolyte. By integrating these parameters, CNN-2 is better equipped to model the complex interactions that occur during dendritic growth, leading to improved prediction accuracy and robustness compared to CNN-1. These additions make CNN-2 more adaptable to varying battery conditions and more reliable in real-world applications.

To provide a clearer comparison of the various calculation methods and models used in the study, we summarize the advantages, disadvantages, applicability, and accuracy (R) of each method in Table \ref{tab:calculation_methods}. This includes classical physics-based approaches such as Molecular Dynamics (MD), Density Functional Theory (DFT), Phase Field (PF), and Monte Carlo (MC), as well as machine learning models like Convolutional Neural Network (CNN-1) and its improved version (CNN-2).

MD and DFT, both of which are widely used in simulating atomic-level interactions, offer high accuracy but are computationally expensive and are typically limited by the available computational power. These methods are well-suited for small systems and for obtaining detailed electronic properties, but their applicability is limited to systems with relatively well-defined atomic structures.

PF and MC models, on the other hand, are particularly effective for simulating complex morphologies and phase transitions, making them ideal for studying dendritic growth mechanisms. While PF excels in handling dynamic processes and complex geometries, it may oversimplify atomic-level details. MC, useful for large-scale statistical modeling, is generally limited by its reliance on random sampling and inability to capture dynamic time evolution precisely.

In contrast, CNN-based models (CNN-1 and CNN-2) provide an efficient and scalable approach to predict dendritic growth patterns, especially in systems with large datasets. CNN-1, while fast and efficient, can struggle with capturing intricate atomic-level interactions, making it less accurate than CNN-2, which benefits from a more complex architecture and better generalization capabilities. As shown in Table \ref{tab:calculation_methods}, CNN-2 achieves higher accuracy (R $\approx$ 0.32–0.38) compared to CNN-1 (R $\approx$ 0.80–0.93), making it a more reliable model for predicting dendritic growth in complex systems.

By providing this comparison, we aim to give readers a comprehensive understanding of the trade-offs between different methods and their performance in predicting the growth mechanisms of battery dendrites. These insights also highlight the potential for combining machine learning with traditional physics-based approaches to enhance the accuracy and efficiency of modeling dendritic growth.

Both VASP calculations (left panels of Figure~\ref{VASPmode}) and the CNN Network-2 model (right panels of Figure~\ref{VASPmode}) were employed for these evaluations. The outcomes offer valuable insights into how the models respond to variations in key parameters. 

The sensitivity analyses systematically manipulated crucial input parameters (t$_{\textrm{i}-1}$: VASP model and t$_{\textrm{i}}$: CNN Network-2 model), encompassing physics-specific attributes such as temperature, pressure, and substrate characteristics. This approach enabled the visualization of the models' responses under diverse conditions and their adaptability to different growth environments.

To assess the sensitivity of the seven physical parameters, we employed a systematic approach where each parameter was varied individually within a predefined range, while keeping the others constant. This allows us to isolate the influence of each parameter on the dendritic growth process. The sensitivity analysis was conducted by running a series of simulations with the CNN-2 model, where the variation of each parameter was observed to determine its impact on the key output variables, such as dendrite length, density, and branching.

Mathematically, the sensitivity of a parameter \( P_i \) can be quantified by calculating the relative change in an output variable \( O \), which is given by the equation:

\[
S(P_i) = \frac{\Delta O}{O_0} \times \frac{P_i}{\Delta P_i}
\]

\( \Delta O \) is the change in the output variable when \( P_i \) is varied,
\( O_0 \) is the initial value of the output variable,
\( P_i \) is the initial value of the physical parameter being varied,
\( \Delta P_i \) is the change in the physical parameter \( P_i \).

This formula calculates the relative sensitivity of the output variable to each parameter. A higher value of \( S(P_i) \) indicates that the output variable is more sensitive to the changes in \( P_i \).

In our analysis, we evaluated how each parameter affects the growth dynamics of dendrites by varying its value within a specified range, such as temperature, from \( T_0 \) to \( T_0 + \Delta T \), and observing the resulting changes in dendrite morphology. For example, when varying the temperature \( T \), we calculated the corresponding changes in dendrite length \( L \), and dendrite density \( D \), as follows:

\[
S(T) = \frac{\Delta L}{L_0} \times \frac{T}{\Delta T}, \quad S(T) = \frac{\Delta D}{D_0} \times \frac{T}{\Delta T}
\]

Where \( L_0 \) and \( D_0 \) are the initial values of dendrite length and density, respectively.

Each parameter was tested for its individual effect on the dendritic growth, and a sensitivity ranking was generated to identify which parameters had the most significant impact on the output variables. This analysis allows us to isolate the most influential factors in dendrite formation and provides insight into how each parameter contributes to the overall electrochemical behavior of the system.

Subsequently, significant insights into the behavior of the CNN Network-2 model were unveiled through sensitivity analysis (Figure~\ref{VASPmode}, right panel). Firstly, the model exhibited resilience to temperature fluctuations, maintaining consistent accuracy across varying temperature conditions, underscoring its efficacy in predicting dendrite growth patterns. Additionally, the model's sensitivity to changes in pressure provided nuanced perspectives on pressure-related dynamics. Changes in substrate properties highlighted the model's responsiveness to surface interactions, crucial for comprehending dendrite growth influenced by substrate composition. Moreover, the sensitivity analysis offered a comprehensive comprehension of the CNN model's behavior, emphasizing its adaptability and shedding light on potential areas for enhancement. These insights underscore the model's reliability across diverse growth scenarios, underscoring its potential for broader application in dendrite growth studies.

\begin{figure*} [t]
\centering \includegraphics[width=0.82\textwidth]{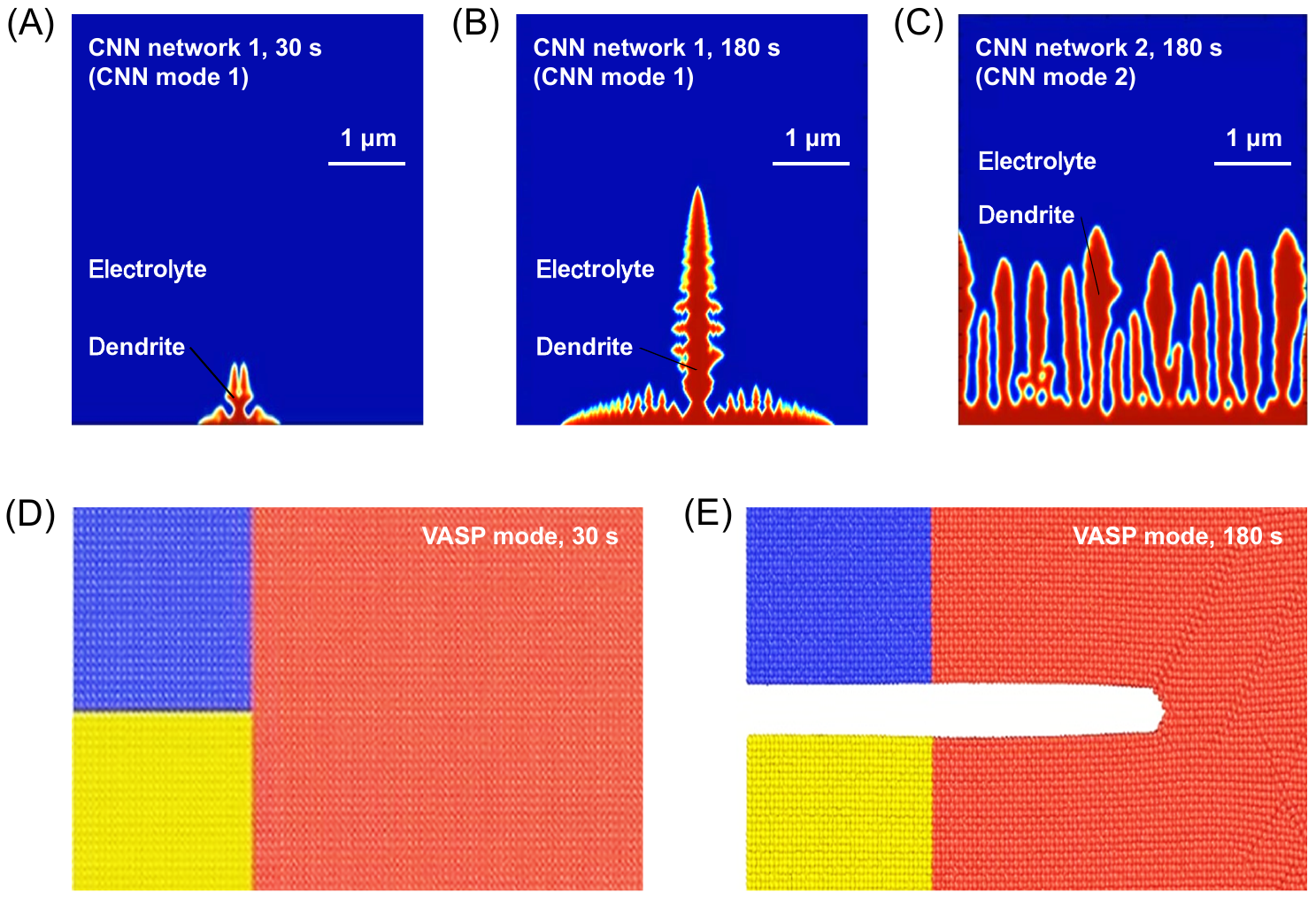}
\caption{Visualization prediction of dendritic growth modes generated by our trained CNN model and VASP calculations. Visualization of dendritic growth prediction by the CNN model at various time points. The red areas indicate dendrite growth, while the blue areas signify the electrolyte. This figure elucidates the model's forecasting of dendritic structure evolution over time, emphasizing alterations in the dendrite growth regions. (A), (B), and (C) illustrate predictions of dendritic growths on the anode of batteries at 30 s, 180 s (generated by CNN network 1), and 180 s (generated by CNN network 2), respectively. (D) and (E) Dendritic growth predictions at 30 s (D) and 180 s (E) simulated by the VASP software, where the white areas depict the dendrites, with the white sticks illustrating the structure and extent of dendrite growth.}
\label{visualization}
\end{figure*}

To validate the predictive capacity of our model, Figure~\ref{visualization} illustrates visual predictions of dendritic growth on battery anodes at various time points. Specifically, Figure~\ref{visualization}(A-C) showcases outputs from our data-driven artificial convolutional neural network model, while Figure~\ref{visualization}(D) and (E) display simulation outputs from the VASP software. It is evident that our trained CNN Network-2 model, representing CNN mode 2, yields the most precise simulation of the dendritic growth process on battery anodes. In these representations, red areas denote dendrites, and blue areas represent the electrolyte. Our model generates insightful visualizations of dendritic growth at different time intervals, offering valuable predictive capabilities for battery performance. Moreover, a supplementary movie (supplemental movie 1) demonstrates the dendritic growth process over a duration ranging from 30 s to 200 s utilizing the CNN Network-2 model. Additionally, supplemental movie 2 illustrates the corresponding dendritic growth process over the same duration but conducted using our artificial CNN Network 1 for comparison. In these visualizations, blue regions signify dendrites, while red regions depict the electrolyte. The visual juxtaposition underscores the CNN Network-2 model's ability to accurately capture intricate growth dynamics, showcasing a close alignment between predicted structures and real-world experimental outcomes. The qualitative analysis further corroborates the quantitative metrics, emphasizing the reliability and accuracy of our CNN Network-2 model in simulating dendritic growth phenomena. These results underscore the significance of incorporating both literature-based and simulated datasets in model training, enhancing the model's adaptability to a broad array of conditions. The achieved levels of accuracy and visual congruence with experimental data establish our CNN model as a valuable instrument for advancing the comprehension and prediction of dendritic growth across diverse scenarios. 
In contrast to CNN-1, CNN-2 incorporates the impact of physical variables on dendritic growth phenomena. By incorporating seven carefully selected physical parameters as inputs, CNN-2 is engineered to more accurately capture the fundamental physical processes. This integration of physical insights bolsters the model's resilience, enabling CNN-2 to offer more dependable and precise forecasts of dendritic growth rates. The inclusion of these parameters aids in unraveling the intricate interactions that govern dendritic formation, rendering CNN-2 a more potent tool for scrutinizing and forecasting dendritic growth in battery systems. 

Our study introduces several innovative aspects in our CNN Network-2 model. While prior studies have delved into data-driven methodologies, we are the pioneers in employing a CNN model and integrating physical parameters as inputs. In contrast to conventional purely data-driven models and solely theoretical computational models,\cite{yan2018computational, akolkar2014modeling} our model showcases improved robustness, providing a more authentic representation of dendritic growth phenomena. These findings highlight the potential of machine learning models in enriching our understanding and forecasting of battery behavior, crucial for the progression of more efficient and dependable energy storage systems.

\subsection{Research Outlook: Practical Applications of CNN-2 in Battery Design}

While the primary focus of this study has been on the accuracy and stability of the CNN-2 model for predicting dendritic growth, it is important to explore how the model can be applied in real-world battery design. Understanding dendrite formation and growth is critical for enhancing the performance, safety, and longevity of batteries, and CNN-2 offers a promising tool for optimizing various aspects of battery design, including electrolyte formulations and electrode materials.

\subsubsection{Optimizing Electrolyte Formulation}
One of the most significant challenges in battery development is minimizing dendrite formation, which can lead to short circuits and performance degradation. CNN-2 can be used to model the effects of different electrolyte compositions on dendritic growth. By varying parameters such as ion concentration, salt composition, and the presence of specific additives, the model can predict how these changes will affect the rate and pattern of dendrite growth under various conditions.

For instance, CNN-2 could simulate the impact of introducing different electrolyte additives that are known to influence the stability of the electrode-electrolyte interface. By predicting the effects of these additives on dendrite formation, the model can guide the selection of additives that inhibit dendrite growth, thereby enhancing the overall performance and safety of the battery. Furthermore, CNN-2 can be used to optimize ion concentration and temperature within the electrolyte to create the ideal conditions for preventing dendrite formation while maintaining efficient battery operation.

\subsubsection{Guiding Electrode Material Design}
CNN-2 also offers valuable insights into electrode material design by simulating how different material properties influence dendrite growth. Electrode materials with specific characteristics—such as surface morphology, porosity, and the presence of protective coatings—can have a significant impact on the formation and growth of dendrites. By incorporating these material properties into the CNN-2 model, it is possible to predict how changes in the electrode's composition or surface treatment could affect the growth of dendrites and the stability of the battery.

For example, the model could be used to assess the effect of electrode coatings that reduce the likelihood of dendrite penetration. CNN-2 could predict how different coating materials or thicknesses influence dendrite growth patterns, helping researchers design more stable and durable electrodes. Moreover, CNN-2 can aid in identifying new electrode materials with better resistance to dendrite formation, thus accelerating the discovery of next-generation materials for advanced batteries.

\subsubsection{Virtual Testing and Accelerated Design Cycles}
The ability to virtually test different electrolyte formulations and electrode materials before conducting physical experiments is a major advantage of using CNN-2 in battery design. By simulating a wide range of conditions and materials, researchers can quickly identify promising candidates for further experimental validation. This not only saves time but also reduces the cost of traditional trial-and-error experimentation. The insights gained from CNN-2 can guide the experimental process, providing a more targeted approach to battery development and potentially leading to safer and more efficient battery systems.

By connecting theoretical predictions with real-world applications, CNN-2 offers a powerful tool for optimizing battery performance and advancing battery technologies. Future work will involve further refinement of the model and the integration of additional physical parameters, which will enhance its ability to predict dendritic growth behavior under a broader range of conditions and lead to more informed design decisions.

\section{CONCLUSION}

In summary, our investigation employed a 2D artificial convolutional neural network-2 model to predict dendritic growth on battery anodes, offering a fresh perspective on comprehending battery performance. Two networks were developed, with network 1 exhibiting limitations in accurately forecasting dendritic growth due to its neglect of various physical variables. Conversely, network 2, incorporating specific physical parameters, demonstrated exceptional accuracy in both training and testing datasets, highlighting its efficacy in modeling dendritic growth phenomena. Additionally, a comprehensive sensitivity analysis was conducted using both the CNN model and VASP software, emphasizing the adaptability and reliability of the CNN Network-2 model across diverse growth scenarios. Insights from this analysis elucidated the model's response to variations in crucial parameters such as temperature, pressure, and substrate characteristics, underscoring its potential for broader applications in dendritic growth research. Visual predictions generated by our CNN Network-2 model exhibited remarkable accuracy, closely mirroring experimental outcomes and offering valuable insights into dendritic growth dynamics. The qualitative analysis complemented the quantitative metrics, affirming the reliability and effectiveness of our model in simulating dendritic growth phenomena. Moreover, our study introduced innovative elements by integrating physical parameters into the CNN model, thereby enhancing its robustness and realism compared to traditional purely data-driven or theoretical computational models. The novelty of the proposed CNN model lies in its ability to integrate various battery components and physical parameters, such as electrolyte concentration, current density, and temperature, to predict dendritic growth in batteries. This comprehensive approach enables the model to capture the complex interactions driving dendrite formation, distinguishing it from traditional models that often focus on limited variables. By incorporating these multiple factors, the model provides a more accurate and detailed prediction of dendritic growth patterns. Furthermore, the model is highly adaptable to specific battery systems. By inputting the relevant physical parameters of a particular battery configuration, the model can predict dendritic behavior, making it suitable for a wide range of battery chemistries and applications. These findings underscore the potential of machine learning models in advancing our understanding and prediction of battery behavior, crucial for the development of more efficient and reliable energy storage systems.

\medskip
\textbf{ACKNOWLEDGEMENTS} \par 

This work was supported by the Science and Technology Development Fund, Macao SAR (File Nos. 0090/2021/A2 and 0104/2024/AFJ), University of Macau (MYRG-GRG2024-00158-IAPME), and the Guangdong{-}Hong Kong{-}Macao Joint Laboratory for Neutron Scattering Science and Technology (Grant No. 2019B121205003).


\medskip
\textbf{CONFLICT OF INTEREST STATEMENT} \par 
The authors declare no conflict of interest.

\medskip
\textbf{DATA AVAILABILITY STATEMENT} \par 
The data that support the findings of this study are available from the corresponding author upon reasonable request.

\medskip

%

%
\clearpage
\begin{table}[ht]
\centering
\begin{tabular}{p{2.5cm}p{3.5cm}p{3.5cm}p{3.5cm}p{2.5cm}}
\hline
\textbf{Method} & \textbf{Advantages} & \textbf{Disadvantages} & \textbf{Applicability} & \textbf{Accuracy (R)} \\ 
MD (Molecular Dynamics) & Can simulate atomic-level interactions & Computationally expensive, limited time scale & Suitable for systems with well-defined atomic interactions; limited by computational power & \( R \approx 0.85 - 0.95 \) \\ 
DFT (Density Functional Theory) & High accuracy for electronic structure & Computationally expensive for large systems, neglects some interactions & Best for electronic properties, small systems, or specific materials & \( R \approx 0.90 - 0.98 \) \\ 
PF (Phase Field) & Effective for modeling complex morphologies, handles dynamic processes & May oversimplify atomic-level details, requires calibration & Ideal for modeling dendritic growth and phase transitions in materials & \( R \approx 0.80 - 0.90 \) \\ 
MC (Monte Carlo) & Useful for statistical modeling, explores a wide range of configurations & May not capture dynamics over time, dependent on random sampling & Suitable for equilibrium properties and large-scale systems where time evolution is less important & \( R \approx 0.70 - 0.85 \) \\ 
CNN-1 (Convolutional Neural Network) & Fast, efficient in predicting dendritic growth patterns & May lack accuracy in capturing detailed atomic-level interactions & Suitable for large datasets with available training data; may have limited generalization & \( R \approx 0.32 - 0.38 \) \\ 
CNN-2 (Convolutional Neural Network, Improved) & Higher accuracy due to more complex architecture, better generalization & Increased computational complexity, requires more training data & Ideal for more complex systems, offering improved prediction accuracy compared to CNN-1 & \( R \approx 0.80 - 0.93 \) \\ \hline
\end{tabular}
\caption{Summary of calculation methods and CNN models used in the study, including advantages, disadvantages, applicability, and accuracy (R).}
\label{tab:calculation_methods}
\end{table}

\clearpage

\clearpage

\clearpage

\clearpage

\clearpage

\clearpage

\begin{figure}
\textbf{Table of Contents}\\
\medskip
  \includegraphics{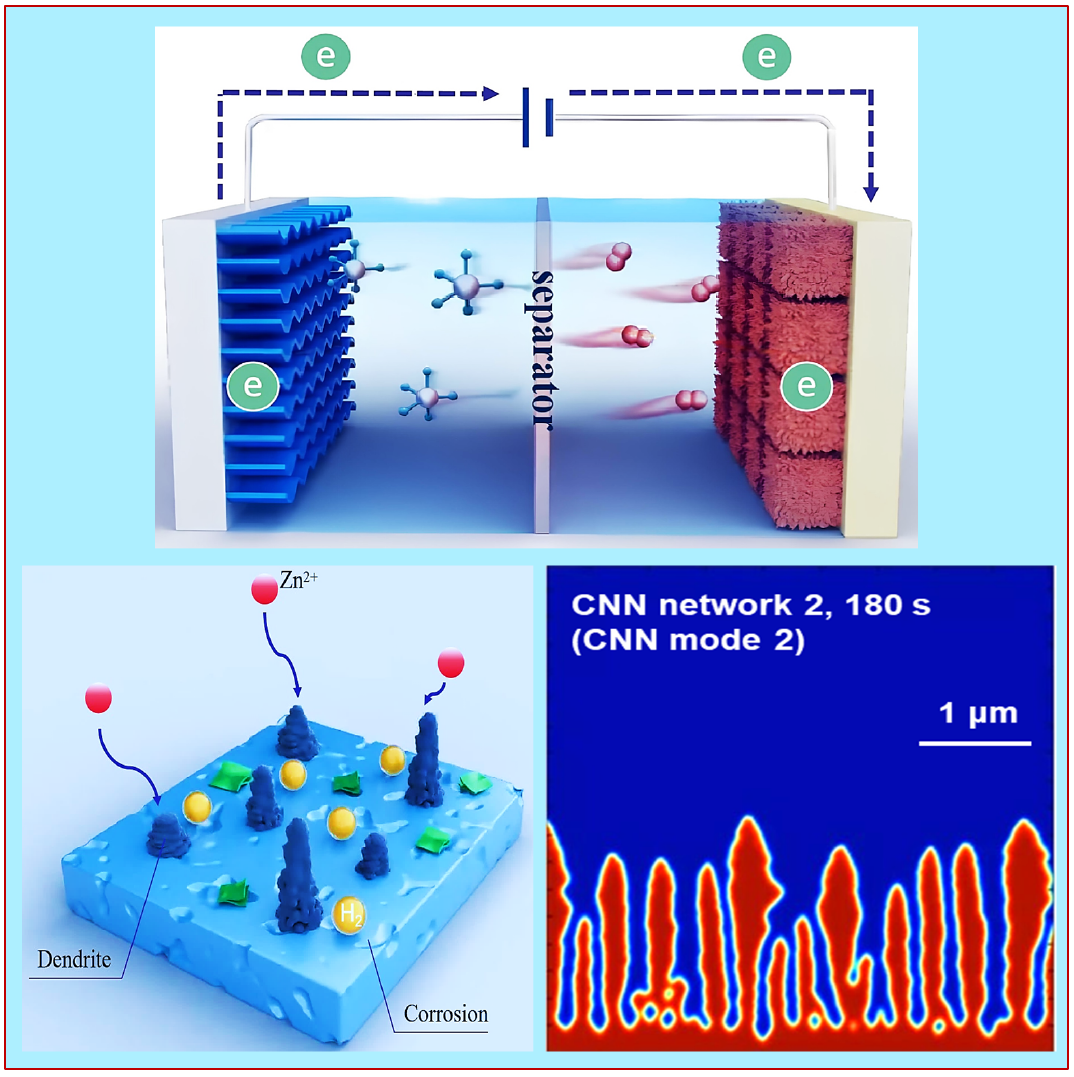}
  \medskip
  \caption*{Dendritic growth in batteries presents challenges to both performance and safety. In this study, we have successfully developed a 2D artificial neural network model that accurately identifies consistent growth modes observed in experimental data. The integration of physical parameters significantly improves accuracy, thereby highlighting the potential of machine learning in predicting battery behavior and advancing energy storage systems.}
\end{figure}

\end{document}